\documentclass[aps,prb,reprint,superscriptaddress,showpacs]{revtex4-1}

\usepackage[utf8]{inputenc}
\usepackage{graphicx}
\usepackage[hmargin=2cm,vmargin=3cm]{geometry}
\usepackage{setspace}
\usepackage{color}
\usepackage{multirow}

\usepackage{amsmath}
\usepackage{amsfonts}
\usepackage{amssymb}
\usepackage[normalem]{ulem}
\usepackage{cancel}
\usepackage{float}
\usepackage{xcolor}
\usepackage{epstopdf}
\usepackage{setspace}
\usepackage{subfigure}

\newcommand{\im}{\mathrm{i}}
\newcommand{\E}{{\bf E}}
\newcommand{\HH}{{\bf H}}

\newcommand{\e}{\varepsilon}
\newcommand{\vPsi}{{\bf\Psi}}
\newcommand{\DP}{k_0\sqrt{\ex}}

\providecommand{\ex}[0]{\varepsilon_{x}} %
\providecommand{\ey}[0]{\varepsilon_{y}} %
\providecommand{\ez}[0]{\varepsilon_{z}} %
\begin{document}

\title{Epsilon-Near-Zero behavior from plasmonic Dirac point: theory and realization using two-dimensional materials}
\author{Marios Mattheakis}
\email[]{mariosmat@seas.harvard.edu}
\homepage[]{http://scholar.harvard.edu/marios$\_$matthaiakis/home}
\affiliation{School of Engineering and  Applied Sciences, Harvard University, Cambridge, Massachusetts 02138, USA}
\affiliation{Crete Center for Quantum Complexity and Nanotechnology,  Department of Physics, University of Crete,  Heraklion 71003, Greece}

\author{Constantinos A. Valagiannopoulos}
\email[]{konstantinos.valagiannopoulos@nu.edu.kz}
\homepage[]{https://sst.nu.edu.kz/konstantinos-valagiannopoulos/}
\affiliation{Department of Physics, Nazarbayev University, Astana 010000, Kazakhstan}
\author{ Efthimios Kaxiras}
\affiliation{School of Engineering and  Applied Sciences, Harvard University, Cambridge, Massachusetts 02138, USA}
\affiliation{Department of Physics, Harvard University, Cambridge, Massachusetts 02138, USA}

\date{\today}

\begin{abstract}
The electromagnetic response of a two-dimensional metal embedded in 
a periodic array of a dielectric host can give rise to a plasmonic Dirac point 
that emulates Epsilon-Near-Zero (ENZ) behavior.  This theoretical result 
is extremely sensitive to structural features like periodicity of 
the dielectric medium and thickness imperfections.  We propose that such a device
can actually be realized by using graphene as the 2D metal and  
materials like the layered semiconducting  transition-metal dichalcogenides or hexagonal boron nitride as the dielectric host.   We propose a systematic approach,  in terms of design characteristics, for constructing metamaterials with   linear, elliptical and hyperbolic dispersion relations which  produce ENZ behavior, normal  or negative diffraction.
\end{abstract}

\maketitle

The confinement of metallic (``free'') electrons in two-dimensional interfaces
can produce powerful effects used to 
drive electromagnetic (EM) devices like nano-antennas with extremely 
short wavelength resonance \cite{Nat424_2003, NanoLet10_2010_antenna}, 
meta-lenses and optical holography \cite{NatComm4_2013, NanoLet14_2014, NanoLet15_2015}, 
active plasmonic systems \cite{PRA89_2014, ieeePhot22_2015, ieee22_2016_costas}, 
and sub-wavelength Bloch oscillations \cite{ JOSAB27_2010, NatCom5_2014}. 
The key feature for such applications is the creation of waves 
propagating along a metal-dielectric interface
with wavelength that is shorter than 
that of the incident radiation, 
while the waves decay exponentially 
in the perpendicular direction.
This surface effect involves electronic motion (plasmons) 
coupled with electromagnetic waves (polariton) and is referred to as Surface Plasmon 
Polariton (SPP). By combining 
the properties of different materials, it is even possible to produce behavior  
not found under normal circumstances like negative refraction \cite{PRL109_2012, APL103_2013, PRB87_2013, MyJAP116_2014, MyNJP16_2014, Nat522_2015}, Epsilon-Near-Zero (ENZ) \cite{PRL97_2006,PRB75_2007,JoNano7_2013},
discrete solitons \cite{PRL99_2007, PRB91_2015solitons} 
and quantum control of light \cite{NatPhys9_2013, NanoTech10_2015}. 

The bottleneck in creating SPP devices with any desirable characteristic 
has been the limitations of typical three-dimensional solids in producing 
perfect interfaces for the confinement of electrons and the features of 
dielectric host. This may no longer be a critical issue.
The advent of truly two-dimensional (2D) materials like graphene (a metal),
transition metal dichalcogenides (TMDC's,  semiconductors)  and hexagonal boron nitride 
(hBN, an insulator) make it possible to produce structures with atomic-level control of features in 
the direction perpendicular to the stacked layers 
\cite{ACSNano6_2012, ACS12_2013,NatCom6_2015, Rodrick}. 
This is ushering a new era in manipulating the properties of 
plasmons and designing devices with extraordinary behavior. 
In particular, 2D structures support 
plasmons (collective excitations) which are fundamentally different from SPPs, 
since the charge carriers  are restricted in two dimensions 
\cite{PRB80_2009, NatPhot6_2012_review}. Nevertheless, 2D plasmons and SPPs 
share similarities in field profiles and in dispersion behavior, 
and could be used interchangeably for the purposes of the present discussion.   
Plasmons in 2D materials exhibit ultra-subwavelength behavior
\cite{PRB80_2009,NatPhot6_2012_review, OP22_2014, JAP115_2014}. 
Graphene is quite special, possessing 
exceptional optical properties due to its high 
quantum efficiency for light-matter interaction \cite{NatPhot6_2012_review, Nat487_2012}. 
Doped graphene has been used as an effective plasmonic platform, since
it supports both high- and low-energy plasmons due
to inter- and intra-band transitions \cite{JModB27_2013}.

ENZ metamaterials exhibit interesting properties like  EM wave propagation 
with no phase delay \cite{PRB75_2007}.  
As a consequence, the pattern of the transmitted/reflected waves can be tailored at will.
Moreover, in a waveguide  filled with ENZ medium all the modes propagate 
irrespective of how small or thin the structure is, exhibiting super-coupling effects \cite{PRL97_2006}.  
Much effort has been devoted toward the design of ENZ media \cite{NJOP15_2013}. 
Here, we propose a systematic method for constructing ENZ metamaterials 
by appropriate combination on 2D materials.
We show analytically that multilayers of a plasmonic 2D material embedded in a dielectric host 
exhibit a plasmonic Dirac point (PDP), 
namely a point in wavenumber space where two linear coexisting dispersion curves 
cross each other, which, in turn, leads to an effective ENZ behavior \cite{PRB91_2015}. 
Specifically, EM wave propagation through layered heterostructures
can be tuned dynamically by controlling the operating 
frequency and the doping level of the 2D metallic layers \cite{PRB87_2013}.
The presence of the PDP is extremely sensitive to structural  features and can only be realized by truly 2D materials, due to the flatness on the atomic-scale that 2D materials provide.  
To prove the feasibility of this design, we investigate numerically 
EM wave propagation in periodic plasmonic structures. They are consisting of 2D metallic layers lying on $yz$ plane in the form of graphene arranged periodically along $x$ axis and possessing surface conductivity $\sigma_s$. The layers are embedded in a uniaxial dielectric host in the form of TMDC or hBN multilayers of thickness $d$
and with non-local relative permittivity tensor $[\e_d]$ with diagonal components $\ex\neq\ey=\ez$. 
We explore the resulting  linear, elliptical and hyperbolic EM dispersion relations which 
produce  ENZ effect, ordinary diffraction and negative
diffraction, depending on the design features.

\begin{figure}[ht!]
\centering
{\includegraphics[scale=.15]{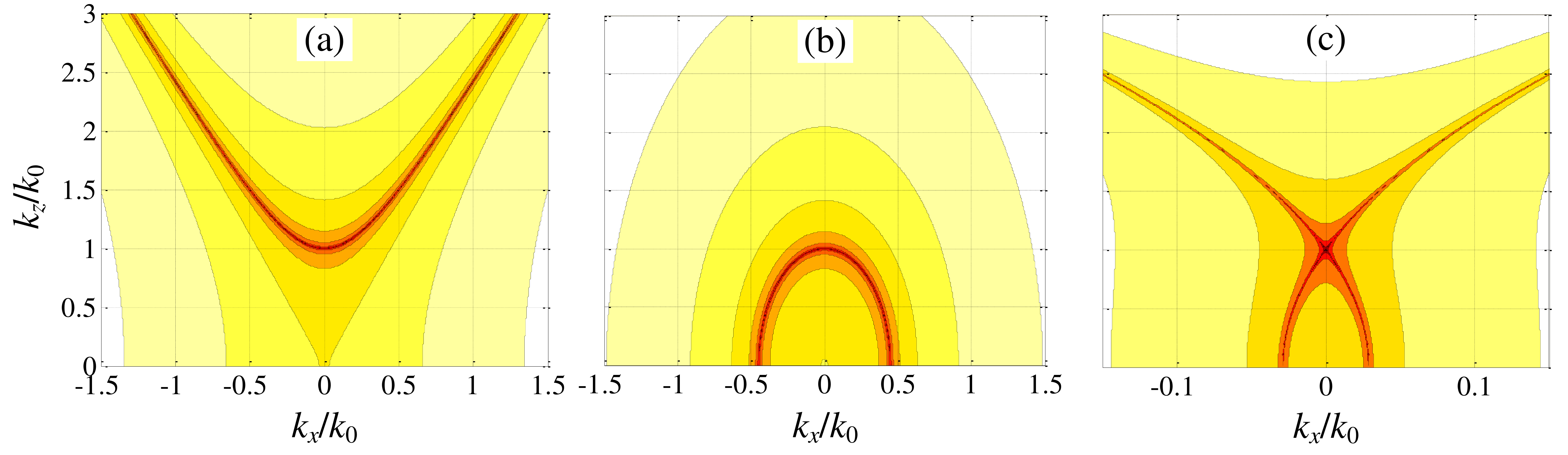}}
\caption{Magnitude of the function $\log|F(k_x,k_z)|$ 
(darker to lighter colors =  smaller to larger values); 
$F(k_x,k_z)=0$ corresponds to black. 
(a) $d<\xi$, strong SPP coupling;
(b) $d>\xi$, weak SPP coupling;
(c) $d=\xi$, plasmonic Dirac point.}
\label{fig:Figs2}
\end{figure}

We solve the analytical problem under transverse magnetic (TM) polarization, 
with the magnetic field parallel to the $y$ direction
which implies that there is no interaction of the electric field with $\varepsilon_y$. 
We consider a lossless host, namely $\ex,\ez\in\mathbb{R}$, 
which is also magnetically inert (relative permeability $\mu=1$).
For  monochromatic harmonic  waves in time with  TM polarization, 
$\E=(E_x,0,E_z)$ and $\HH=(0,H_y,0)$, Maxwell equations lead to the three equations 
connecting the components of the $\E$ and $\HH$ fields.
For the longitudinal component \cite{PRL109_2012, PRL99_2007}, 
$E_z=(\im\eta_0/k_0\ez)(\partial H_y/\partial x)$
where $k_0=\omega/c$ is the vacuum wavenumber at frequency $\omega$
and $\eta_0=\sqrt{\mu_0/\varepsilon_0}$ is the free space impedance.
Defining the vector of the transversal field components as 
$\vPsi=\left(E_x ~~H_y\right)^T$, gives \cite{PRL99_2007} 
\begin{equation}
\label{eq:FullMatrix}
\im\frac{\partial}{\partial z}\vPsi= k_0\eta_0\left({\begin{array}{cc} 0    & 1+\frac{1}{k_0^2}\frac{\partial}{\partial_x}\frac{1}{\ez}\frac{\partial}{\partial_x}\\ 
               \frac{\ex}{\eta_0^2} & 0\      \end{array}} \right) \vPsi
\end{equation}
For EM waves propagating along the $z$ axis, namely $\vPsi(x,z)= \vPsi(x) e^{\im k_z z}$, 
we obtain the eigenvalue problem for  
the wavenumber $k_z$ of the  SPPs along $z$
\cite{PRL109_2012, PRL99_2007}.
The metallic 2D planes carry a surface current $J_s=\sigma_s E_z$, which acts
as a boundary condition in the eigenvalue problem.   
The magnetic field must be $H_y^-(x) e^{\im k_z z}$ for $-d<x<0$ 
and $H_y^+(x) e^{\im k_z z}$ for $0<x<d$ on either side of the metallic plane at $x=0$, 
with boundary conditions
 $H_y^+(0)-H_y^-(0)=\sigma_s E_z(0)$ and $\partial_xH_y^+(0)=\partial_xH_y^-(0)$.
Using the Bloch character along $x$, due to the periodicity of the system, with Bloch wavenumber $k_x$: 
$H_y^+(x)=H_y^-(x-d)e^{\im k_x d}$, 
we arrive at the dispersion 
relation \cite{PRL109_2012}:
\begin{equation}
\label{eq:dispersion}
F(k_x,k_z) = 
\cos(k_x d) - \cosh(\kappa d) + \frac{\xi\kappa}{2}\sinh(\kappa d) = 0
\end{equation}
where $\kappa^2=(\ez/\ex)(k_z^2-k_0^2\ex)$ 
expresses the anisotropy of the  host medium and 
$\xi=-(\im\sigma_s\eta_0/k_0\ez)$ 
is the ``plasmonic thickness'' 
which determines the SPP decay length \cite{PRL109_2012, JAP115_2014}. 
For lossless 2D metallic planes, $\sigma_s$ is purely imaginary and
$\xi$ is purely real (from the assumption of $\ez\in\mathbb{R}$). 
At the center of the first Brillouin zone $k_x=0$, the equation has a trivial solution \cite{PRL109_2012} for $\kappa=0 \Rightarrow k_z=\DP$
which corresponds to propagation of $x$-polarized fields 
travelling into the host medium with refractive index $\sqrt{\e_x}$ 
without interacting with the 2D planes which are positioned along $z$ axis \cite{MyNJP16_2014}. 
Near the Brillouin zone center $(k_x/k_0\ll 1$ and $\kappa \simeq 0)$  
and under the reasonable assumption of a very dense grid $(d\rightarrow 0)$, 
we obtain $k_x d\ll 1$  and  $\kappa d\ll 1$, 
we Taylor expand Eq. (\ref{eq:dispersion}) to second order in $d$:
\begin{equation}
\label{eq:taylor1}
\frac{k_z^2}{\ex}+\frac{d}{(d-\xi)\ez}k_x^2=k_0^2.
\end{equation}
From a metamaterial point of view \cite{PRB87_2013,MyJAP116_2014}, the entire
system is treated as a homogeneous anisotropic medium
with effective relative permittivities  given by
\begin{equation}
\label{eq:effperms}
\e^{\rm eff}_{x}=\e_x, \; \;  
\e^{\rm eff}_{z}=\e_z+\im \frac{\eta_0 \sigma_s}{k_0d}=\e_z\frac{d-\xi}{d}.
\end{equation}
The approximate dispersion relation Eq. (\ref{eq:taylor1})  
is identical to that of an equivalent homogenized medium described by 
Eq. (\ref{eq:effperms}): $k_z^2/\e^{\rm eff}_{x}+k_x^2/\e^{\rm eff}_{z}=k_0^2$. 
Indeed, a very dense mesh of 2D media is a prerequisite for homogenization \cite{OP22_2014, Photonics12_2014,prb90_2014}. 
Eq. (\ref{eq:effperms}) indicates the capability to control the behavior of the overall structure along the $z$ direction:
the choice $d=\varepsilon_z/(\varepsilon_z-\varepsilon_x)\xi$
leads to an isotropic effective medium, $\e^{\rm eff}_{z}= \e^{\rm eff}_{x}$. 
For the lossless case, $\xi\in\mathbb{R}$, we identify three possibilities, provided
an ordinary material ($\ex,\ez>0$) is used as host:\\
(i) $\xi>d$, strong SPP coupling: SPPs  develop along the $z$ direction at the interface between 
the conducting planes and the dielectric host.
In this case, the overall effective response of 
the system becomes also plasmonic, 
with Bloch plasmon polaritons waves \cite{prb90_2014} created along the $x$ direction. 
The shape of the supported band on the $(k_x,k_z)$ plane is hyperbolic, since the system behaves as a hyperbolic metamaterial \cite{PRL109_2012,  MyNJP16_2014, Photonics12_2014} with $\e^{\rm eff}_{z}<0$, 
$\e^{\rm eff}_{x}>0$, Fig. \ref{fig:Figs2}(a) .\\
(ii) $0<\xi<d$, weak SPP coupling: since $\xi$ is still positive, SPPs develop along the $z$ direction 
between the conducting plane and the dielectric host. 
However, the effective behavior of the entire structure is not dominated by SPP coupling \cite{PRL109_2012} and the shape of the dispersion relation on the $(k_x,k_z)$ plane is an ellipse since $\e^{\rm eff}_{z}, \e^{\rm eff}_{x}>0$, Fig. \ref{fig:Figs2}(b).\\
(iii) $\xi<0$:   
in this case, the  2D planes  do not support plasmonic modes. 
The dispersion relation on the $(k_x,k_z)$ plane is an ellipse, 
as in an ordinary photonic crystal \cite{Photonics12_2014,APL103_2013}, with  $\e^{\rm eff}_{z}, \e^{\rm eff}_{x}>0$, Fig. \ref{fig:Figs2}(b).\\
When either the 2D medium (${\rm Re}[\sigma_s]\ne 0$) 
or the host material (${\rm Im}[\e_z]\ne 0$) are lossy, 
a similar separation holds by replacing $\xi$ by ${\rm Re}[\xi]$. 

The most interesting case is the linear dispersion, 
where $k_z$ is linearly dependent on $k_x$ and $dk_x/dk_z$ is constant for a wide range of 
$k_z$ \cite{PRL109_2012, Photonics12_2014}.  When this condition holds, 
the spatial harmonics travel with the same group velocity into the effective medium 
 \cite{PRL109_2012, APL103_2013}. 
To engineer our structure to exhibit a close-to-linear dispersion relation, 
we inspect the approximate version of Eq. (\ref{eq:taylor1}):
a huge coefficient for $k_x$ will make $k_0^2$ on the right hand side 
insignificant; if $\xi=d$, the term proportional to $k_x^2$ 
increases without bound yielding a linear relation between $k_z$ and $k_x$.
With this choice, $\sigma_s=-\im(k_0d\ez/\eta_0)$, 
and substituting in the exact dispersion relation Eq. (\ref{eq:dispersion}), 
we find that $(k_x,k_z)=(0,k_0\sqrt{\ex})$ becomes a saddle point for the transcendental function $F(k_x,k_z)$ giving rise to the conditions for the appearance of two permitted bands, namely two lines on the $(k_x,k_z)$ plane across which $F(k_x,k_z)=0$. 
This argument connects a mathematical feature, 
the saddle point of the dispersion relation, with a physical feature, 
the crossing point of the two coexisting linear  dispersion curves, 
the Plasmonic Dirac point \cite{PRB91_2015} 
as indicated in Fig. \ref{fig:Figs2}(c). 
From a macroscopic point of view, the choice $\xi=d$ makes 
the effective permittivity along the $z$ direction vanish, as is evident from
Eq. (\ref{eq:effperms}). The existence of a  PDP makes 
the effective medium behave like an ENZ material in one direction ($\e^{\rm eff}_{z}=0$). 

The very special behavior associated with the PDP would be of restricted usefulness 
if its existence were sensitively depended on the the exact fulfillment of the condition  Re$[\xi]=d$, where losses have been taken into account (Im$[\xi]\neq0$).
For this reason, we investigate the behavior of the gap created between the two bands on the $(k_x,k_z)$ plane, when the PDP breaks down.
At the center of the Brillouin zone $(k_x=0)$, 
where the minimum gap is created \cite{PRB91_2015, Photonics12_2014, prb90_2014}, 
the dispersion relation Eq. (\ref{eq:dispersion}) is: 
$g(k_z)=2 (\cosh(\kappa d)-1)/ (\kappa \sinh(\kappa d))=\xi$.
Near the PDP, $k_z=k_0\sqrt{\ex}+\Delta k_z$ and $\xi=d+\Delta \xi$, while
$g(k_z)\cong g(k_0\sqrt{\ex})+g^{\prime}(k_0\sqrt{\ex})\Delta k_z$;
given that $g(k_z)=d$
and $g^{\prime}(k_z)=-(k_0 d^3 \ez/6\sqrt{\ex})$
in the limit $k_z\rightarrow k_0\sqrt{\ex}$, 
a direct relation between $\Delta k_z$ and $\Delta \xi$ is obtained
\begin{equation}
\label{eq:gap}
\frac{\Delta k_z}{k_0\sqrt{\ex}}=-\frac{6}{(k_0d)^2\ez}\frac{\Delta \xi}{d}.
\end{equation}
In the derivation of Eq. (\ref{eq:gap}), 
we assume that $\Delta \xi$ is small compared to $d$ in the vicinity of the PDP, that is, 
$d={\rm Re}[\xi]$, leading to  two conditions:  ${\rm Re}[\Delta \xi]/d \ll 1$ 
and ${\rm Im}[\Delta \xi]/d \ll1$. 
Near the PDP the former condition is satisfied, 
since ${\rm Re}[\Delta \xi] = {\rm Re}[\xi]-d \rightarrow 0$. 
For the imaginary part we have, 
${\rm Im}[\Delta \xi] /d={\rm Im}[\xi]/ {\rm Re}[\xi]={\rm Re}[\sigma_s]/ {\rm Im}[\sigma_s]\ll 1,$
which is satisfied if the system is characterized by  low losses. 
To illustrate the situation with an example, 
we use the Drude model to describe the conductivity of a 2D metal, as is appropriate for  doped graphene. 
In this case  ${{\rm Re}[\sigma_s]}/{{\rm Im}[\sigma_s]}={1}/{\tau \omega},$ where $\tau$ accounts for losses. For representative values of $\tau$ and $\omega$ 
we obtain $1/(\tau \omega) \simeq 10^{-2}$, which makes our assumption 
of low losses  reasonable. 
Moreover, in Eq. (\ref{eq:gap})  the real and imaginary  parts have been decoupled, that is, the losses, corresponding to ${\rm Im}[\Delta \xi]$, do not affect the band-gap given by 
${\rm Re}[\Delta k_z]$.

The choice $k_z=k_0\sqrt{\ex}$ works as a trivial solution of the dispersion equation regardless of the values of the rest of the parameters. 
Consequently, in the vicinity of $\xi=d$, Eq. (\ref{eq:gap}) gives 
the relative spread of the gap $(\Delta k_z/k_0\sqrt{\ex})$ between the two bands 
since it implies that $k_z=k_0\sqrt{\ex}+\Delta k_z$ is also a solution of 
Eq. (\ref{eq:dispersion}) at $k_x=0$. 
Since the lattice of the 2D medium is electrically dense ($k_0d\ll 1$), 
Eq. (\ref{eq:gap}) indicates a substantial sensitivity of the PDP on the value of $\Delta\xi$. 
As a consequence, a small error on  the $\xi=d$ condition leads to a significant gap between 
the two bands: taking an isotropic silica glass with $\ex=\ez=4$ as host material, 
a deviation of order 
$(\Delta\xi/d)\cong 10\%$ gives rise to a band gap of 
order $\Delta k_z\cong 10^3 k_0$ for $k_0 d=10^{-2}$.
It should be additionally stressed that only one band moves from 
the PDP position: for $\Delta\xi<0\Rightarrow\xi<d$ 
the upper point of the elliptical band remains at $(k_x,k_z)=(0, k_0\sqrt{\ex})$, 
whereas the hyperbolic band
moves to higher values of $k_z$ at a rate given by Eq. (\ref{eq:gap})
with the converse behavior for $\Delta\xi>0\Rightarrow\xi>d$.

\begin{figure}[ht!]
\centering 
\subfigure{\includegraphics[scale=0.2]{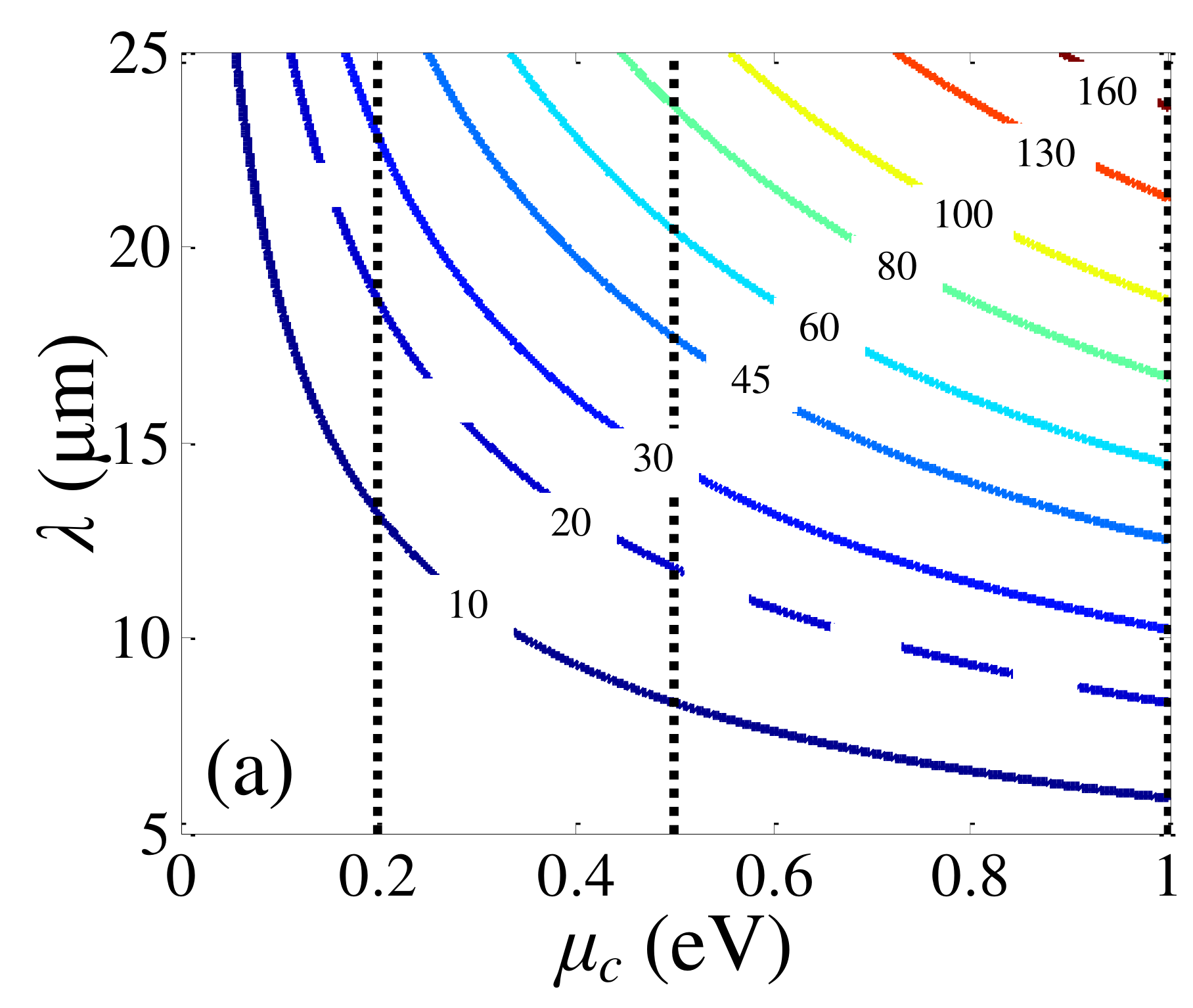} 
\label{fig:Fig3a}}
~~~\subfigure{\includegraphics[scale=0.2]{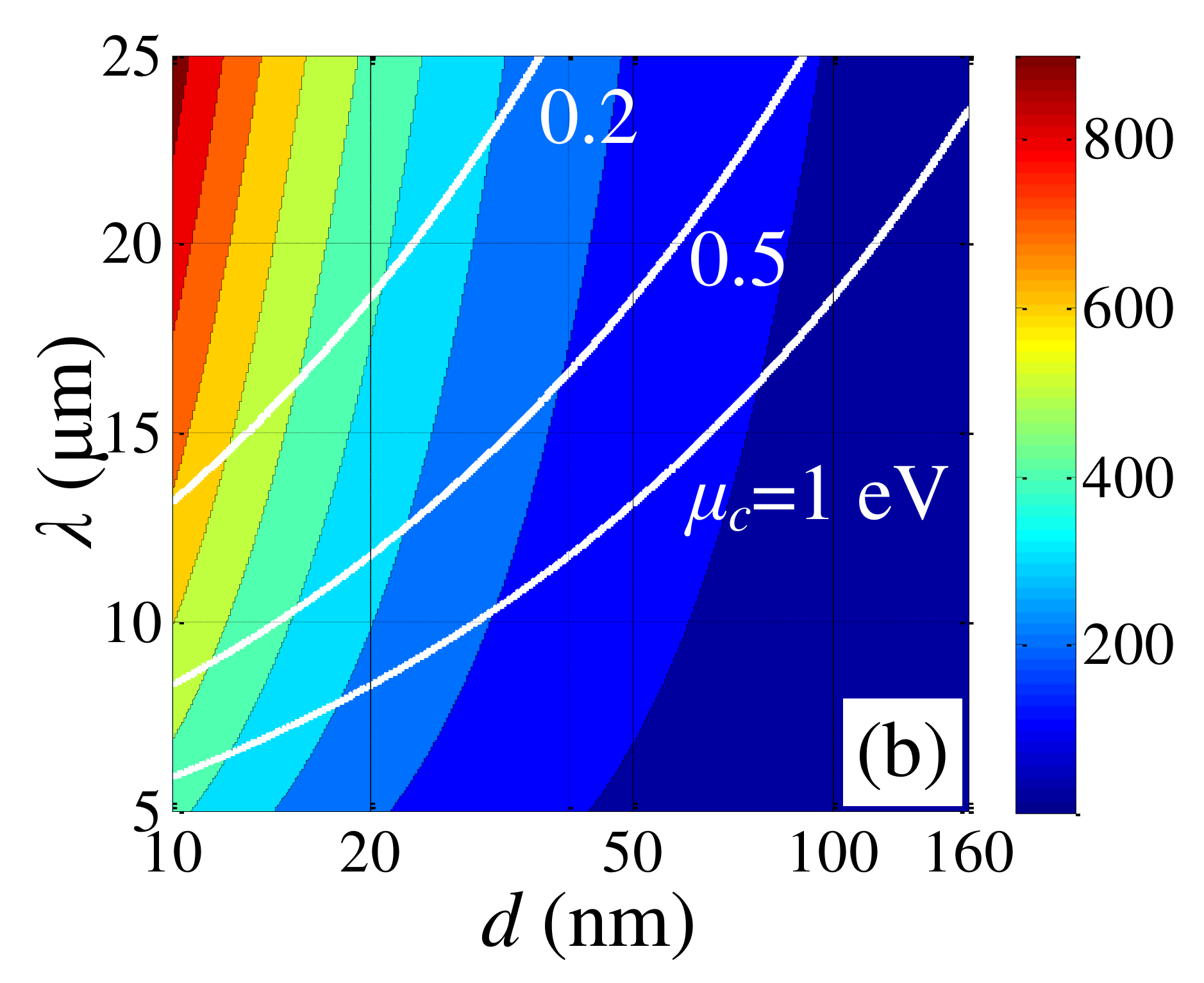}  
\label{fig:Fig3b}}
\subfigure{\includegraphics[scale=0.2]{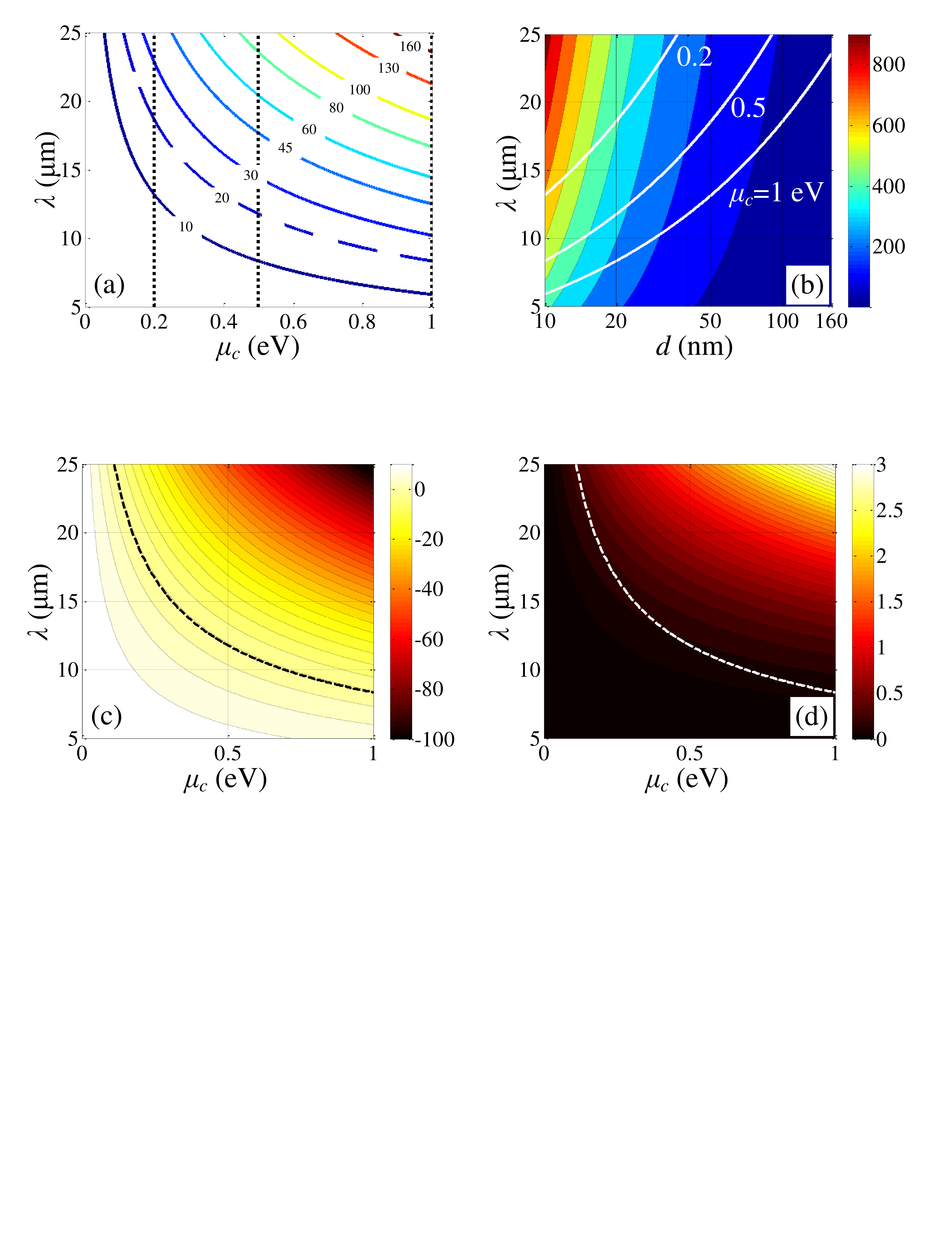}  
\label{fig:Fig3c}}
\subfigure{\includegraphics[scale=0.2]{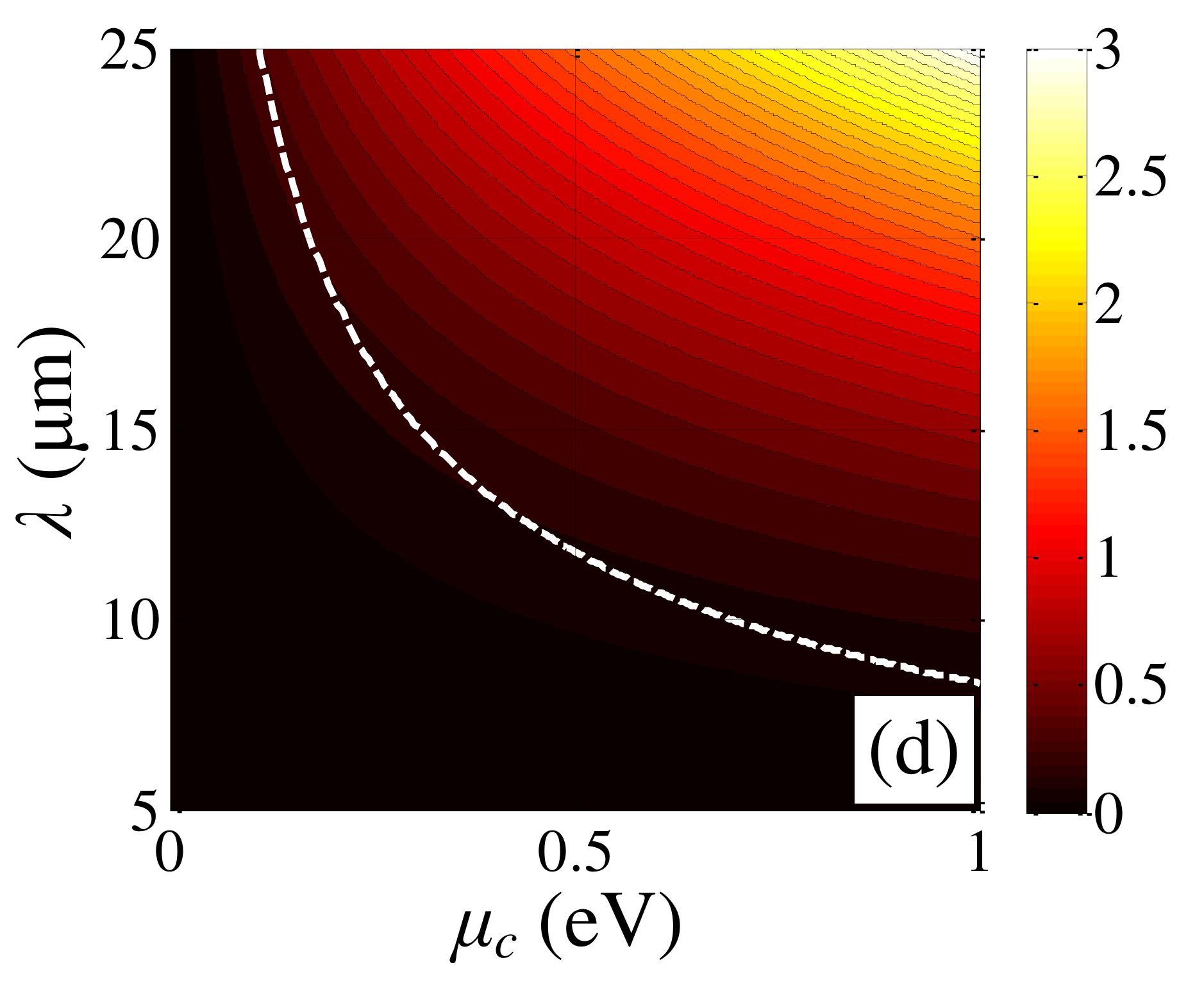}  
\label{fig:Fig3d}}
\caption{(a) Combinations of graphene chemical potential $\mu_c$ 
and free space operational wavelengths $\lambda$ 
leading to ENZ behavior (PDP in dispersion relation)  
for several lattice periods $d$ (in $\rm{nm}$). 
(b) The propagation  distance $L$ of a plasmonic mode in units of $d$ 
for all the combinations of the wavelength $\lambda$ and the period $d$ leading to ENZ effect. 
The white lines show representative levels of graphene doping [dotted lines in (a)]. 
(c) Real (d) imaginary parts of the effective permittivity $\e_z^{\rm eff}$ 
of the effective medium  for the choice $d=20~\rm nm$
[dashed line in (a)]; dashed curves indicate the ENZ regime.}
\label{fig:Figs3}
\end{figure}

The extreme sensitivity of the PDP on the spatial period $d$ 
between the 2D planes makes the use of regular materials as  dielectric
hosts impractical,  unless the dielectric host is also a 2D material with 
atomic scale control of the thickness $d$ and no roughness.
For instance, one could build 
the dielectric host by stacking 2D layers of materials 
like hBN \cite{NanoTech10_2015, ACSNano6_2012} or molybdenum disulfide (MoS$_2$)  \cite{ACS12_2013,  NatCom6_2015,  Rodrick} with essentially perfect planarity, 
complementing the planarity of graphene, which has been used extensively in 
optoelectronic and plasmonic applications \cite{ NatPhot6_2012_review}. 
The surface conductivity $\sigma_s$ of an infinite graphene plane 
includes both intraband and interband transition contributions \cite{JModB27_2013}, 
with the intraband contribution  dominating at THz frequencies
which is approximated by the Drude model, 
$\sigma_s(\omega)=\im e^2 \mu_c/[\pi\hbar^2(\omega+\im/\tau)]$, 
where $\mu_c$ is the tunable chemical potential and $\tau$ is the transport 
scattering time of the electrons \cite{NanoLet14_2014,PRL109_2012, NatPhot6_2012_review, Photonics12_2014}.
In the following, we use bulk MoS$_2$, which at THz frequencies 
is assumed lossless with a diagonal permittivity tensor of elements,
$\ex\cong 3.5$ (out of plane) 
and $\ey=\ez\cong13$ (in plane) \cite{ACS12_2013, NatCom6_2015,  Rodrick}. 
The optical losses of graphene are taken into account using $\tau=0.5$ ps \cite{PRL109_2012}.  
In Fig. \ref{fig:Fig3a}, we show the combinations of $\mu_c$  
and the operational wavelength in free space  $\lambda$ which lead to a PDP 
 for several values of lattice density distances $d={\rm Re}[\xi]$ in $\rm nm$.

The ENZ behavior should be accompanied by low effective losses,
otherwise the propagating field is damped fast.  
A crucial quantity demonstrating the efficiency of the proposed medium is 
the length $L$ that an EM wave can propagate into such a device without 
losing a significant part of its power.  We find that the length $L$ before 
the amplitude falls to the $1/e$ of its maximal value, in units of the period $d$, is given by:
\begin{equation}
\label{eq:Lod}
\frac{L}{d}=\sqrt{\frac{2}{\e_z}}\sqrt{\frac{{\rm Im}[\sigma_s]}{{\rm Re}[\sigma_s]}}\frac{1}{k_0d}.
\end{equation}
From Eq. (\ref{eq:Lod}), the propagating beam travels along $x$ for more lattice periods, 
the less lossy the graphene sheets and the denser the lattice. 
The seeming contradiction of longer propagation in a denser lattice 
can be explained by the stronger SPP coupling for smaller periods $d$ \cite{PRL109_2012}. 
Using the  Drude model for $\sigma_s$ gives $L/d=\sqrt{c\tau \lambda}/(d\sqrt{\e_z\pi})$. 
This is shown in Fig. \ref{fig:Fig3b} by a contour plot as a function of  
free-space wavelength $\lambda$ and cell physical size $d$, and takes values in the range 
several hundreds.  The loss-unaffected transmission length in terms of the number of wavelength cannot be determined explicitly by Eq.  (\ref{eq:Lod}), because $\sigma_s$ depends on $\lambda$. Nevertheless, an explicit expression of $L$ in terms of the number of wavelengths 
$\lambda$ can be calculated, in the context of the Drude model.
Eq.  (\ref{eq:Lod}) can be re-written as  $L/\lambda=\sqrt{c\tau/(\varepsilon_z\pi \lambda})$ indicating that  $L/\lambda$ is inversely proportional to $\sqrt{\lambda}$ and has no dependence on the doping level $\mu_c$. The proposed design exhibits no significant losses of the propagating wave 
even in long structures consisting of several hundred periods. 
Interestingly, the best results (smallest losses) do not require large graphene doping.

To illustrate, for a reasonable 
distance between successive graphene planes of $d=20~\rm nm$, 
the real (Fig. \ref{fig:Fig3c}) and imaginary (Fig. \ref{fig:Fig3d}) 
permittivity values that can be emulated by this specific graphene-MoS$_2$ 
architecture determine the device characteristics at different frequencies ($\lambda$) 
and graphene doping levels ($\mu_c$).
Positive values of ${\rm Re}[\e^{\rm eff}_{z}]$ 
are relatively moderate and occur 
for larger frequencies and lower doping levels of graphene; on the other hand,
${\rm Im}[\e^{\rm eff}_{z}]$ is relatively small in the ENZ region 
as indicated by a dashed line in both graphs. Such a fact renders our theoretical assumptions for 
lossless structures quite realistic; however, losses become larger as ${\rm Re}[\e^{\rm eff}_{z}]$ 
gets more negative. 
As these results show, the extreme sensitivity of the PDP on $d$ can 
be turned into an advantage for device fabrication: the proper 
combination of $d$ and $\mu_c$ can be selected to produce a device 
that operates at a given frequency ($\lambda$), as the practical limitations of 
layer stacking ($d$) or graphene doping ($\mu_c$) dictate.

\begin{figure}[ht!]
\centering
\subfigure{\includegraphics[scale =0.25]{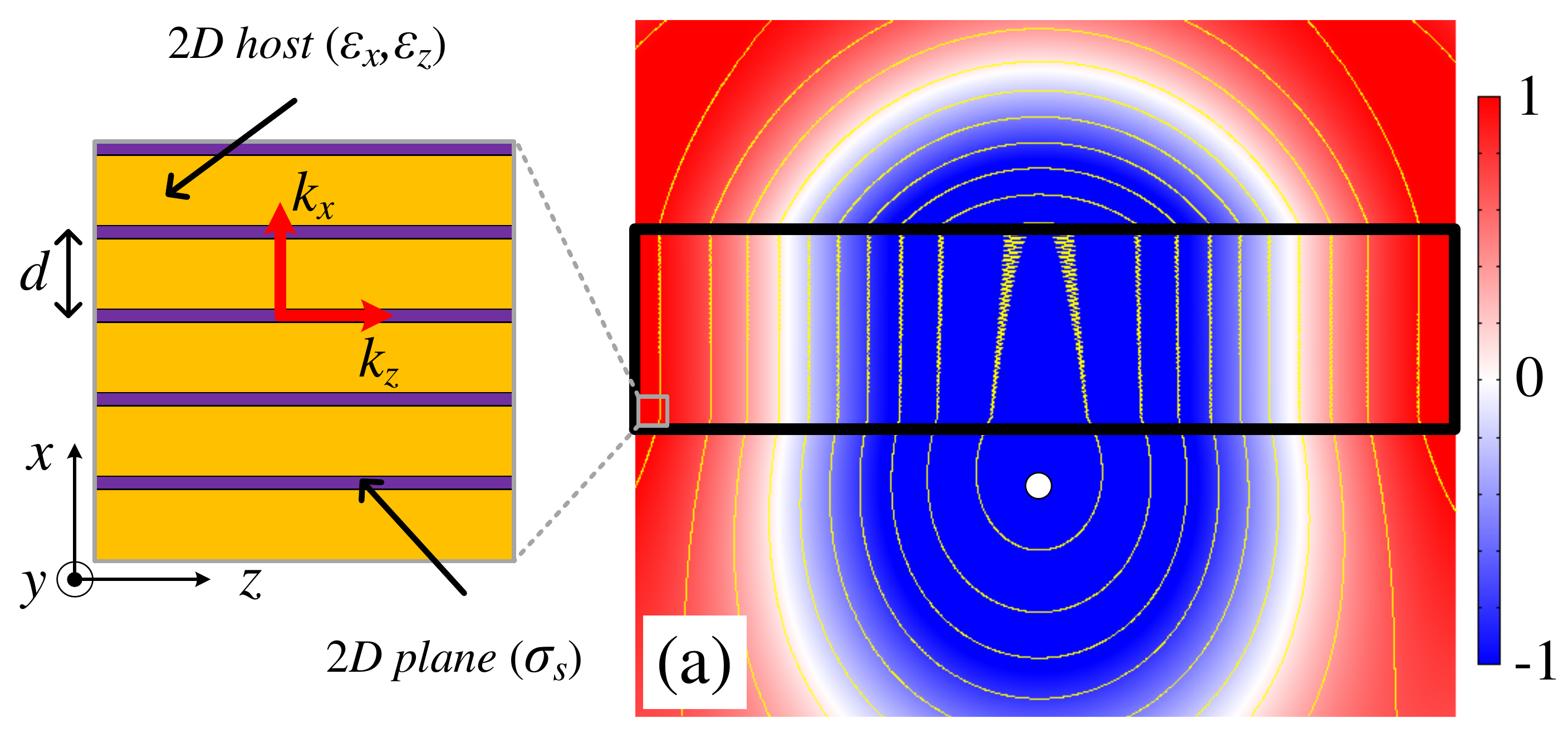}
   \label{fig:enz}} \\
\subfigure{\includegraphics[scale =0.25]{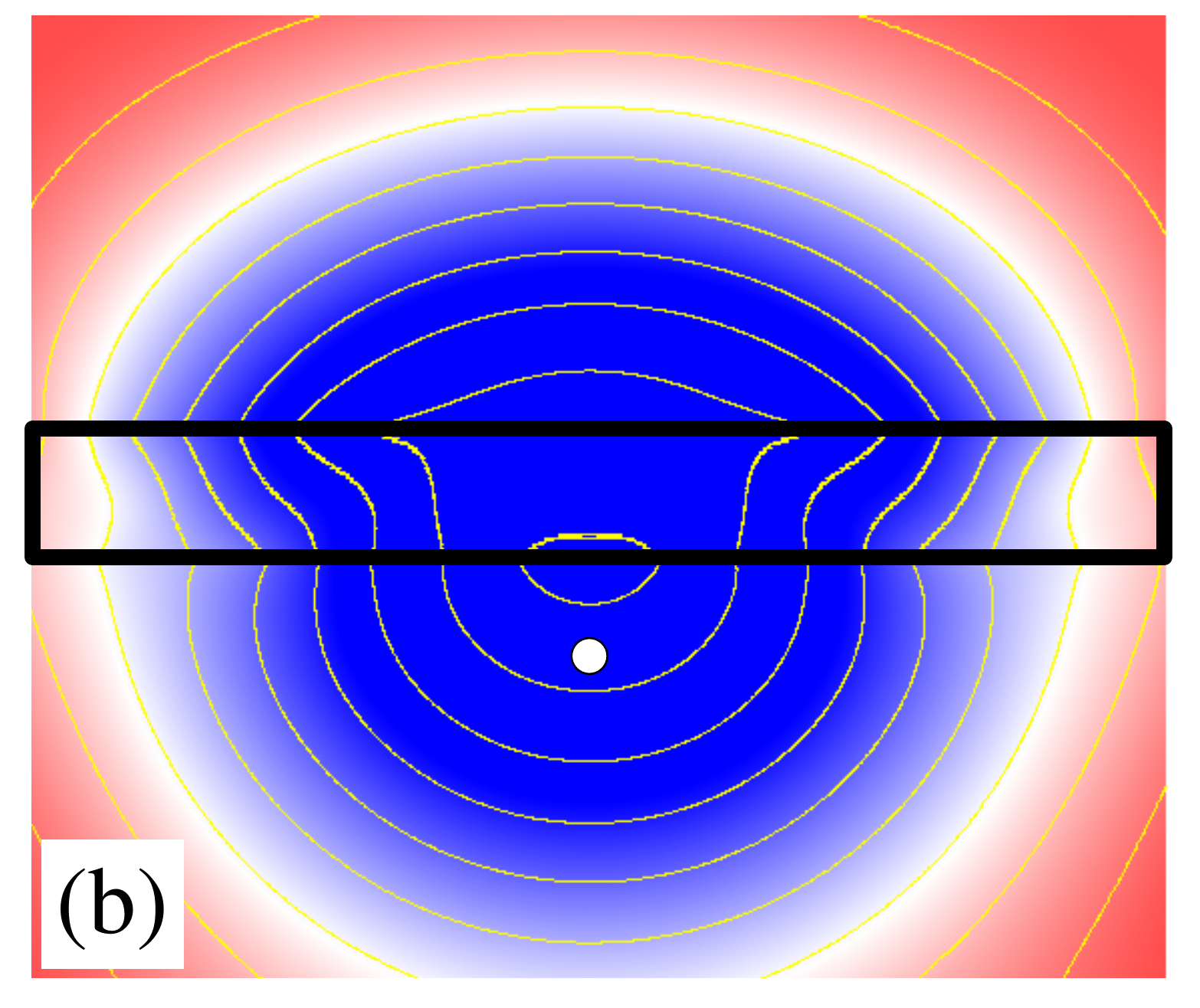}
   \label{fig:hyperbolic}} 
\subfigure{\includegraphics[scale =0.25]{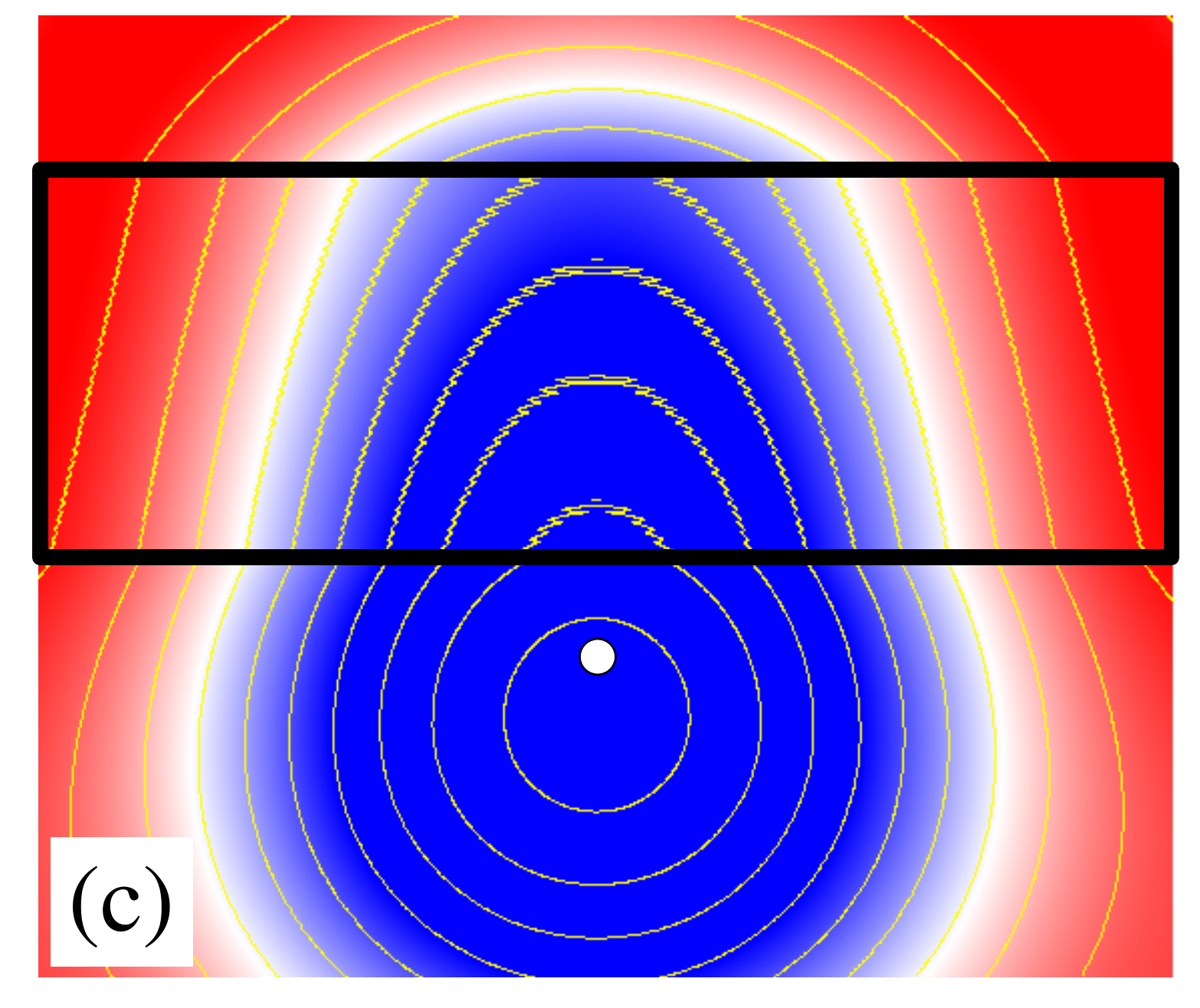}
   \label{fig:elliptic}}
\caption{Spatial distribution of the axial magnetic field of a device consisting of  40 graphene sheets embedded in MoS$_2$ background excited by a magnetic dipole at the point marked as white dot. The thick black boundary defines the volume containing the graphene multilayers. 
(a) $d={\rm Re}[\xi]$ (ENZ behavior); the inset has details of the graphene multilayer configuration. 
(b) $d=0.5{\rm Re}[\xi]$ (hyperbolic metamaterial).
(c) $d=1.5{\rm Re}[\xi]$ (elliptical medium). ${\rm Re}[\xi]=20.8~\rm nm$.}
\label{fig:dispersionNUM}
\end{figure}

To examine the actual EM field distribution in our graphene-MoS$_2$ 
configuration for each of the three 
characteristic cases of supported bands, we excite a finite structure  
consisting of 40 graphene planes and ${\rm Re}[\xi]=20.8~\rm nm$ for operational wavelength 
in vacuum $\lambda=12~\rm \mu m$, 
using as a source a 2D magnetic dipole positioned close to one of its two interfaces 
and oriented parallel to them; this choice of source  allows us to study the system's response 
when exciting all the incidence angles with the same power.
The spatial distribution of the magnetic field value is shown in Fig. \ref{fig:dispersionNUM}
where the volume containing the graphene multilayers is denoted by a thick black frame. In all three cases, 
the reflections are negligible because the background region is filled 
with a medium of the same dielectric properties as MoS$_2$. 
In Fig. \ref{fig:enz}, the system is in the critical case ($d={\rm Re}[\xi]$), where the wave propagates through the graphene sheets without dispersion as in an ENZ medium. 
In Fig. \ref{fig:hyperbolic}, the interlayer distance is $d=0.5{\rm Re}[\xi]$ (strong SPP coupling regime) and the system shows negative (anomalous) diffraction with the front of the propagating wave into the multilayered structure showing a hyperbolic shape. 
In Fig. \ref{fig:elliptic}, $d=1.5{\rm Re}[\xi]$ (weak SPP coupling regime) and the EM wave shows ordinary diffraction through the graphene planes.

We acknowledge discussions with J. D. Joannopoulos, M. Soljacic, 
G.P. Tsironis, S. Shirodkar and P. Cazeaux, and 
support by EFRI 2-DARE NSF Grant 1542807 (MM), 
ARO MURI Award No. W911NF14-0247 (EK),
the E.U. program FP7-REGPOT-2012-2013-1, grant 316165 (MM),  
and a SEED grant of Nazarbayev University for preliminary research (CAV). 
We used computational resources on the Odyssey cluster of the FAS 
Research Computing Group at Harvard University.

\end{document}